\documentclass[a4paper,9pt]{article}
\usepackage{spconf,amsmath,graphicx}
\usepackage{setspace, subfig}
\usepackage{xcolor,url}
\usepackage{multirow}
\usepackage[skip=2pt]{caption}

\title{L2 PROFICIENCY ASSESSMENT USING SELF-SUPERVISED SPEECH REPRESENTATIONS}
\name{Stefano Bannò$^{1,2}$, Kate M. Knill$^{3,4}$, Marco Matassoni$^1$, Vyas Raina$^3$, Mark J. F. Gales$^{3,4}$}
\address{$^1$Fondazione Bruno Kessler, Trento, Italy, 
$^2$University of Trento, Trento, Italy\\
$^3$ALTA Institute, Cambridge University, UK, $^4$Enhanced Speech Technology Ltd., UK\sthanks{This paper reports on research partially supported by Cambridge University Press \& Assessment, a department of The Chancellor, Masters, and Scholars of the University of Cambridge.}}
\begin{document}
\ninept
\maketitle
\begin{abstract}
There has been a growing demand for automated spoken language assessment systems in recent years. A standard pipeline for this process is to start with a speech recognition system and derive features, either hand-crafted or based on deep-learning, that exploit the transcription and audio. Though these approaches can yield high performance systems, they require speech recognition systems that can be used for L2 speakers, and preferably tuned to the specific form of test being deployed. Recently a self-supervised speech representation based scheme, requiring no speech recognition, was proposed. This work extends the initial analysis conducted on this approach to a large scale proficiency test, Linguaskill, that comprises multiple parts, each designed to assess different attributes of a candidate's speaking proficiency. The performance of the self-supervised, wav2vec 2.0, system is compared to a high performance hand-crafted assessment system and a BERT-based text system both of which use speech transcriptions. Though the wav2vec 2.0 based system is found to be sensitive to the nature of the response, it can be configured to yield comparable performance to systems requiring a speech transcription, and yields gains when appropriately combined with standard approaches. 

\end{abstract}
\section{Introduction}

In recent years, the growing number of learners of English as a second language (L2) on a global scale has created an increasing demand for automated spoken language assessment systems for applications in  Computer-Assisted Language Learning (CALL) in private practice and classroom situations and to certify language exams. One of the compelling reasons for automatic assessment is the need to evaluate and provide feedback to increasing numbers of learners and return results in a timely manner. Secondly, compared to human graders, not only can automatic systems ensure greater speed, but they can do it at a lower cost, since the recruitment and training of new human experts is expensive and can offer only a small increase in performance. Finally, the use of automatic assessment methods can enhance consistency, reliability, and objectivity of scoring, since machines are not susceptible to rater effects and - more simply - to tiredness~\cite{vanmoere2017}. \par
Automated systems for assessing L2 speaking proficiency typically receive sequential data from a learner as input to predict their grade or level with respect to overall proficiency or specific facets of proficiency. This input data may consist - as needed - of phones, recognised words, acoustic features, or other information derived directly from audio or from automatic speech recognition (ASR) transcriptions. In most cases, sets of hand-crafted features related to specific aspects of proficiency, such as fluency~\cite{strik1999automatic}, pronunciation \cite{chen}, prosody~\cite{coutinho2016assessing} and text complexity~\cite{bhat2015automatic}, are extracted and fed into graders to predict analytic grades related to those specific aspects. An alternative but similar approach consists of concatenating multiple hand-crafted features related to multiple aspects of proficiency in order to obtain overall feature sets, which are subsequently projected through graders to predict holistic grades, as shown in~\cite{muller2009automatically, crossley2013applications, wang2018towards, liu2020dolphin}. The efficacy of hand-crafted features for scoring either overall proficiency or individual aspects heavily relies on their specific underlying assumptions, and salient information about proficiency may risk being discarded. For holistic grading, this limitation has been addressed by substituting hand-crafted features with automatically derived features for holistic grading prediction, either through end-to-end systems~\cite{chen2018end} or in multiple stages~\cite{cheng2020asr, takai2020deep}. Other studies have investigated the use of graders that are trained on holistic scores but are defined with both their inputs and structure adapted to target particular aspects of proficiency, such as pronunciation~\cite{kyriakopoulos2018deep}, rhythm~\cite{kyriakopoulos2019deep} and text~\cite{wang2021, raina2020universal}. In these cases, a possible limitation might be the absence of information concerning facets of proficiency that are not present in the input data fed to the grader, although we have shown that it is possible to combine multiple graders targeting different aspects of proficiency in a previous study \cite{banno2022view}. Such a limitation is particularly evident in systems using ASR transcriptions, since a) they 
will contain errors and so will not
provide faithful verbatim transcriptions of a learner's performance, thus not rendering its content appropriately; b) 
transcriptions do not contain any information relating to the message realisation,
such as fluency, pronunciation, intonation, rhythm, or prosody\footnote{Some information about pronunciation can be obtained from the performance of the ASR system and associated confidence scores, whereas hesitation markers, truncated words, and other disfluencies (if available from the ASR transcript) might be considered proxies of fluency.}. Instead, they represent a precious resource for highly specific tasks in CALL applications, e.g., for spoken grammatical error correction and feedback, as we have shown in ~\cite{lu2022}. \par
In this paper, in order to tackle these issues and address these limitations, we propose an approach based on self-supervised speech representations using wav2vec 2.0~\cite{baevski}. Recent studies have shown that self-supervised learning (SSL) is an effective approach in various downstream tasks of speech processing applications, such as ASR, keyword spotting, emotion recognition, intent classification, speaker identification, and speaker diarisation~\cite{baevski, yang21c}. In these studies, contextual representations were applied by means of pre-trained models. In particular, it has been demonstrated that these models are able to capture a vast range of speech-related features and linguistic information, such as audio, fluency, suprasegmental pronunciation, and even semantic and syntactic text-based features for L1, L2, read and spontaneous speech~\cite{shah}. In the field of CALL, SSL has been investigated for mispronunciation detection and diagnosis~\cite{wu21, xu21, peng21} and automatic pronunciation assessment~\cite{eesung}.

\begin{figure}[h!]
\includegraphics[scale=0.3]{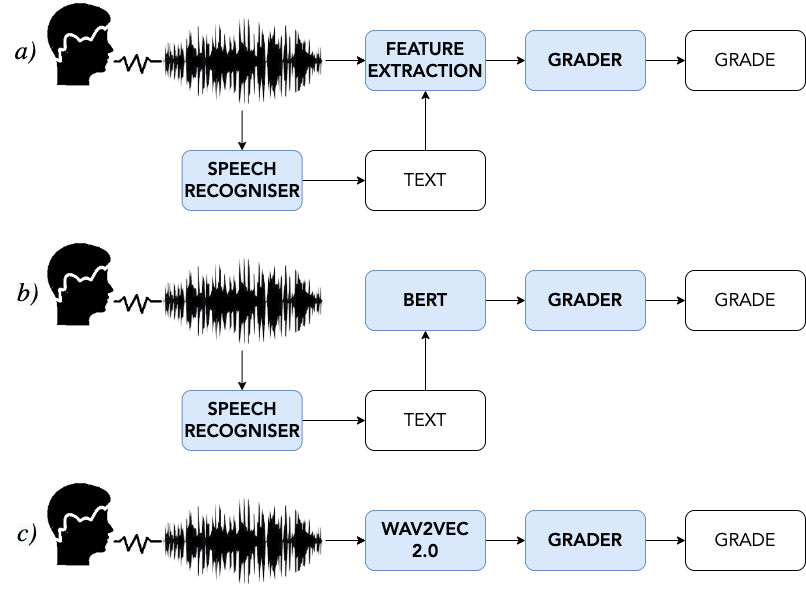}
\caption{The three systems considered in this study: a) standard grader, b) BERT-based grader, and c) wav2vec2-based grader.}
\label{pipelines_compared}
\centering
\end{figure}

To the best of our knowledge, our previous work~\cite{banno2022wav} was the first to propose the use of SSL for holistic and analytic proficiency assessment. Nevertheless, this study had two limitations: a) the relatively small amount of data used in the experiments and b) the comparison with a BERT-based~\cite{devlin2018} baseline only, which, although fed with manual transcriptions containing hesitations, false starts, and truncated words, did not consider purely acoustic features, thus potentially missing strictly speech-related aspects of proficiency. To address these limitations, in the present contribution, we conduct our experiments of proficiency assessment using a large amount of L2 learner data and comparing the performance of wav2vec 2.0 to two types of grader: a BERT-based grader and a standard grader fed with a set of hand-crafted features across different facets of proficiency (see Figure \ref{pipelines_compared}). In addition to this, we test the effectiveness of wav2vec 2.0 on a multi-part examination predicting both the overall grades and the individual grades of each part of the exam. Furthermore, we investigate various combinations between the wav2vec2-based, the BERT-based, and the standard graders.

In Section 2, we describe the data used in our experiments. Section 3 shows the model architectures and the combinations considered in our study. Section 4 illustrates the results of our experiments. Finally, in Section 5, we discuss and analyse the results.

\section{DATA}

The data used in our study are obtained from candidate responses to the spoken parts of the Linguaskill examinations for L2 learners of English, provided by Cambridge English Language Assessment~\cite{ludlow2020official}. The Linguaskill speaking exam is divided into five parts. In Part 1, the candidates should answer eight questions about themselves, of which the first two are not marked. The answers last about 10 or 20 seconds. Part 2 consists of a reading aloud activity including eight sentences of 10 seconds each. Part 3 and Part 4 test the candidates' ability to deliver a long turn, speaking for up to one minute. In the first, the candidates should talk about a given topic, whereas in the latter they should describe one or more graphics, such as charts, diagrams, or information sheets. Finally, in Part 5, test-takers should give their opinions in form of responses of about 20 seconds to five questions related to a given topic. Since each part contributes 20\% to the speaking exam, the overall grade is calculated as the average of the grades assigned to the five parts.

Each speaker is graded on a scale from 1 to 6 based on the proficiency scales of the Common European Framework of Reference (CEFR) for languages~\cite{cefr2001}, i.e., A1, A2, B1, B2, C1, and C2. 

Non-overlapping datasets of 31475 and 1033 speakers are used as the training and development/calibration set, respectively. 
For evaluation, we consider two test sets, LinGen and LinBus, of 1049 and 712 speakers, respectively. LinGen includes learners' responses to questions on General English, while LinBus contains answers to questions on Business English. Each test set features around 30 L1s and is balanced for gender and proficiency level.

The text transcriptions for training and test used by the standard and BERT-based graders are generated using a non-native English ASR system. A Kaldi-based system with a TDNN-F acoustic model and a trigram language model is used similar to that in~\cite{lu2019impact} with average WER of $\sim$20\%.

\section{MODEL ARCHITECTURES}
\noindent
\textbf{Wav2vec2-based graders:} in wav2vec 2.0, speech audio is encoded through a multilayer convolutional neural network (CNN). After encoding, masking is applied to spans of the resulting latent representations. Subsequently, these are fed into a transformer to build contextualised representations. Gumbel softmax is used to calculate the contrastive loss on which the model is trained, and speech representations are learned from this training.

For our experiments, we initialised the model configuration and processor from a version provided by HuggingFace~\cite{huggingface}\footnote{\url{huggingface.co/patrickvonplaten/wav2vec2-base}}.
After the learners' answers are fed into the model, wav2vec 2.0 provides contextualised representations. To handle representations of various audio lengths, we employed a mean pooling method to concatenate 3D representations into 2D representations, which are finally passed through a regression head~\cite{banno2022wav}. Since we trained a grader for each part of the exam, after trying different architectures, for Part 1 and Part 5 we used a regression head composed of a layer of 768 units, a Dropout layer, and the output layer, whereas, for Part 2, Part 3 and Part 4, we used a deeper architecture, consisting of a stack of three layers of 768 units, a Dropout layer, a layer of 128 units, and, finally, the output layer. The graders use mean squared error (MSE) as the loss function. Training uses the AdamW optimiser~\cite{adamw} and hyperparameters vary depending on each part. For Part 1, we used batch size 16, gradient accumulation step 2, dropout rate 0.1, and learning rate 5e-5, and we trained the grader for 2 epochs. For Part 2, we used batch size 16, gradient accumulation step 2, dropout rate 0.5, and learning rate 1e-6, and we trained the grader for 3 epochs. For Part 3 and Part 4, the hyperparameters are the same: we set batch size to 8, gradient accumulation step to 4, dropout rate to 0.5, and learning rate to 1e-5, and we trained the grader for 2 epochs. Finally, the grader for Part 5 has batch size set to 8, gradient accumulation step to 2, dropout rate to 0.1, and learning rate to 5e-5, and we trained it for 1 epoch.

As has been said, the first part of wav2vec 2.0 consists of a stack of CNN layers that are used to obtain acoustically meaningful - but contextually independent - features from the raw speech audio signal. This part of the model has already been sufficiently trained during the pre-training stage and does not need fine-tuning. Therefore, for our experiments we froze the parameters of the feature extractor.

\noindent
\textbf{BERT-based graders:} for comparison, we use the text grader presented in \cite{raina2020universal}, which consists of an LSTM with attention over its hidden representation. The inputs are word embeddings obtained by passing the words of each utterance through a trained BERT language model~\cite{devlin2018}.

\noindent
\textbf{Standard graders:} we also compare our SSL approach to a standard grader, which is a Deep Density Network (DDN) trained on a set of hand-crafted features designed to cover all the different aspects of proficiency and is described in~\cite{malinin2017}. These features include: grade dependent language model and word level statistics; statistics of phone duration; statistics to capture rhythm; fluency metrics; and fundamental frequency statistics are used to represent intonation. More detailed information about the features employed can be found in~\cite{wang2018towards}.

For all graders, their predictions are the result of ensembles. Further information about the ensemble approach can be found in~\cite{xixin2020}.
Systems are calibrated and in a final set of experiments combined using linear combination:
\begin{equation}
    {\hat y}^{(n)} =\beta_0 + \sum_{p\in{\cal P}}\beta_p {\hat y}^{(n)}_p
\end{equation}
where ${\cal P}$ is the set of parts to combine, which may come from multiple systems, and $\beta_p$ are the coefficients associated with the parts. For the baseline submission performance  the values of $\beta_p, p>0$ are all constrained to be the same, and equal to 0.2, to yield simple averaging consistent with the combination of operational examiner scores. Ordinary Least Squares (OLS) estimation using the development/calibration set is used to find the values of $\beta_p$ when unequal weighting is used.

For the evaluation of the grading systems at the per-part level, we use root-mean-square error (RMSE), whilst further comparisons also include Pearson's correlation coefficient (PCC), Spearman's rank correlation coefficient (SRC), and the percentage of the predicted scores that are equal to or lie within 0.5 (i.e., within half a grade) ($\%\leq{}0.5$), and within 1.0 (i.e., one grade) ($\%\leq{}1.0$) of the actual score.

\section{EXPERIMENTAL RESULTS AND ANALYSIS}

\noindent
\textbf{Part-Level Performance:} we start our series of experiments with grading each of the five parts of the exam. For this part of our analysis we only consider LinGen. Table~\ref{T:lingen_rmse} reports the results of the three grading systems in terms of RMSE.

\begin{table}[ht!]

\small
\begin{center}

\begin{tabular}{ l  c  c  c  c  c  c} \hline

 \textbf{Model} & \textbf{P1} & \textbf{P2} & \textbf{P3} & \textbf{P4} & \textbf{P5}  \\ \hline

 std (${\tt s_d}$) & 0.625  & 0.662  & 0.671  & 0.686  & 0.633 \\
 BERT (${\tt b_t}$) & 0.628 & 0.683 & 0.681 & 0.694 & 0.629\\
 wav2vec2 (${\tt w_v}$) & 0.601  & 0.827  & 0.845 & 0.845  & 0.674 \\

\hline

\end{tabular}
\end{center}

\caption{RMSE results on the five parts of the LinGen exam.}
\label{T:lingen_rmse}
\end{table}

We can see that the wav2vec 2.0 performance varies across the parts, with close or better RMSE to the other two graders for Parts 1 and 5 and lower performance on Parts 2, 3 and 5. This 
appears to be due to the nature of the responses required for different parts. Parts 1 and 5 consist of several short spontaneous answers, whereas Parts 3 and 4 are also composed of spontaneous speech but a single longer and more complex response in each case. 
The lower wav2vec 2.0 performance may be due to our use of a mean pooling method, which may be giving too compressed a representation for longer utterances. 

Part 2, by contrast, is similar in length to Part 1 but consists of read speech responses. This part mainly targets pronunciation and fluency skills at the expense of content-related aspects of proficiency so the wav2vec 2.0 system might have been expected to do well. 
Its higher RMSE might be due to the absence of information related to the reference text read by the test-takers, which is present in the other two grading systems. It is noticeable that the standard grader, which covers all aspects of a candidate's speech, performs the best, with the BERT grader which cannot measure pronunciation or prosody slightly behind. 


\noindent
\textbf{Submission-Level Performance:} the second part of our experiments is focused on the overall grades of the Linguaskill exam, i.e., the average of the grades assigned to the five parts. Table \ref{T:summary_results_overall} shows the results of the three grading systems both on LinGen and LinBus.

\begin{table}[ht!]
\small
\begin{center}

\begin{tabular}{ l  c  c  c  c  c  } \hline
\multicolumn{1}{c}{}  &     \multicolumn{5}{c}{\textbf{LinGen}} \\
 \textbf{Model} & \textbf{PCC} & \textbf{SRC} & \textbf{RMSE} & \textbf{\%$\leq$0.5} & \textbf{\%$\leq$1.0}  \\ \hline 

 ${\tt s_d}$ & 0.932   & 0.937   & 0.383   & 81.5   & 98.6  \\
 ${\tt b_t}$ & 0.929  & 0.934  & 0.395  & 80.3  & 98.5   \\
 ${\tt w_v}$ & 0.908   & 0.931   & 0.455   & 73.3   & 97.3   \\

\hline

\end{tabular}
\end{center}

\begin{center}

\begin{tabular}{ l  c  c  c  c  c } \hline
\multicolumn{1}{c}{} &     \multicolumn{5}{c}{\textbf{LinBus}} \\
  \textbf{Model} & \textbf{PCC} & \textbf{SRC} & \textbf{RMSE} & \textbf{\%$\leq$0.5} & \textbf{\%$\leq$1.0}  \\ \hline

 ${\tt s_d}$ & 0.911  & 0.918 & 0.416   & 76.5   & 98.3    \\
 ${\tt b_t}$ & 0.920  & 0.925  & 0.398  & 80.1 & 99.2   \\
 ${\tt w_v}$ & 0.893  & 0.911  & 0.446   & 72.1   & 97.9     \\


\hline

\end{tabular}
\end{center}

\caption{Submission-level performance on LinGen and LinBus.}
\label{T:summary_results_overall}
\end{table}

They all achieve good results across all metrics, with the standard grader and the BERT-based grader performing moderately better than the wav2vec2-based grading system. As regards the BERT-based grader specifically, we have already observed analogous trends in \cite{raina2020universal} and \cite{banno2022view}, and this fact is quite significant, since it highlights the importance of content-related aspects of speaking proficiency, which is far from being a mere surrogate of fluency and pronunciation. Furthermore, it appears that the standard grading system outperforms the other two graders on LinGen, but has a slightly worse performance than the BERT-based grader on LinBus. This might be ascribable to the different language models, since the first test set contains questions on General English, whereas the latter includes questions on Business English, which is typically more specific and complex. 

\begin{figure}
\centering
\begin{subfloat}
   \centering
   {\footnotesize \textbf{std}}
   \includegraphics[width=1\linewidth]{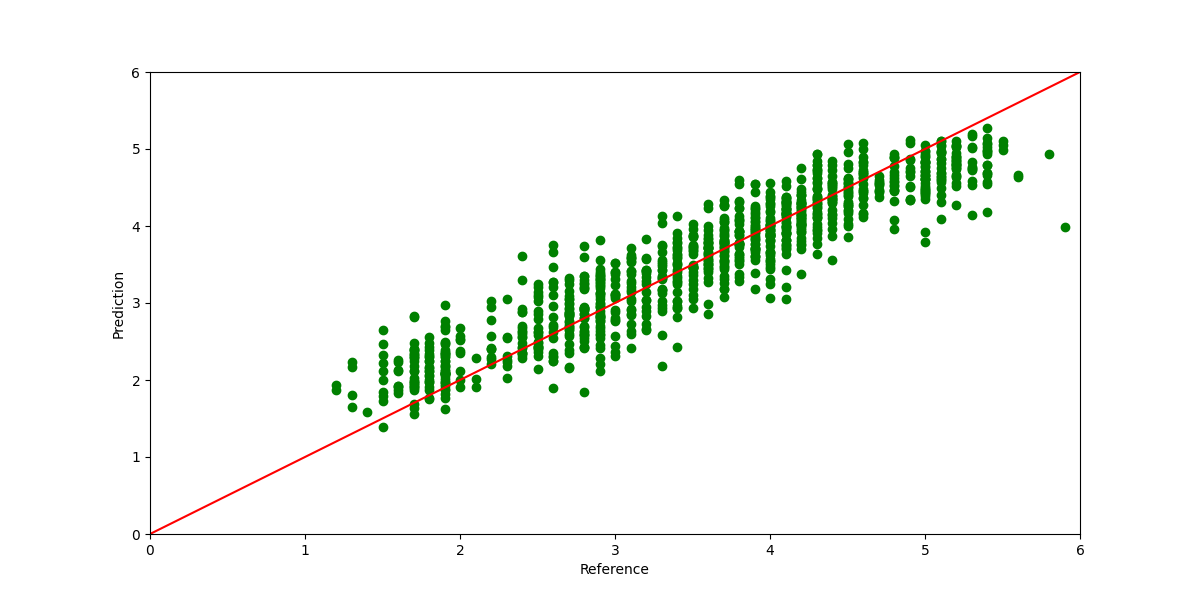}
   \label{fig:dev02_std} 
\end{subfloat}

\begin{subfloat}
   \centering
   {\footnotesize \textbf{wav2vec 2.0}}
   \includegraphics[width=1\linewidth]{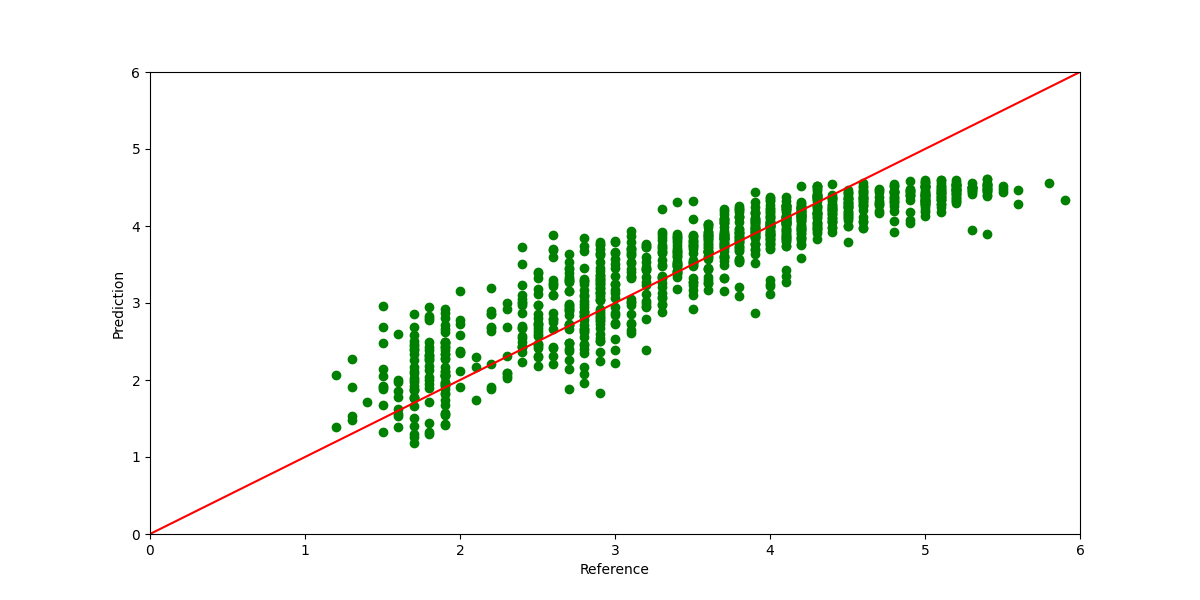}
   \label{fig:dev02_wav}
\end{subfloat}

\caption[]{Reference vs predicted scores for standard and wav2vec2-based graders on LinGen.}
\label{fig:scatter_dev02}
\end{figure}

Figure~\ref{fig:scatter_dev02} shows that the wav2vec2-based graders can discriminate between lower levels of proficiency, but we can clearly see that it is not able to distinguish between the highest levels as its maximum prediction is 4.6, i.e., between grades B2 and C1 (the plots for LinBus show a similar trend).
Our hypothesis is that higher-level assessment tends to be more dependent on message construction (what is said) rather than message realisation (how it is said), and wav2vec 2.0 does not have actual knowledge of words.



\begin{figure}
\centering
\begin{subfloat}
   \centering
   {\footnotesize \textbf{per-part combination ${\tt s_d}$$\otimes$${\tt b_t}$$\otimes$${\tt w_v}$}}
   \includegraphics[width=1\linewidth]{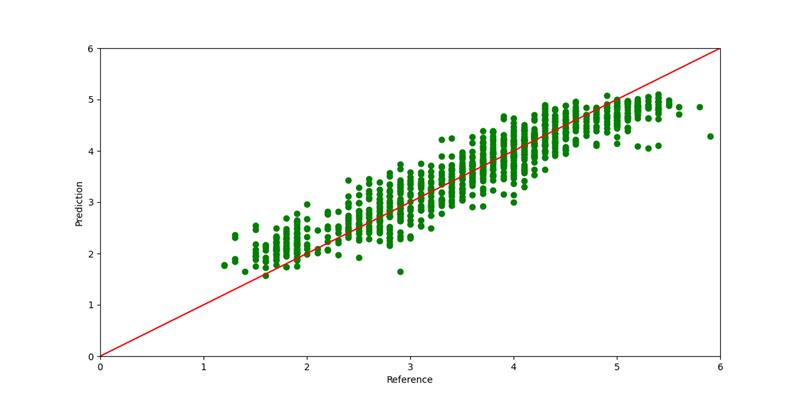}
   \label{fig:dev02_comb} 
\end{subfloat}
\caption[]{Reference vs predicted scores for combined system (${\tt s_d}$$\otimes$${\tt b_t}$$\otimes$${\tt w_v}$)}
\label{fig:scatter_dev02_comb}
\end{figure}

\noindent
\textbf{Combinations:} as a preliminary step, we investigated combinations of the grading systems by calculating their simple average and using a multiple linear regression model fit with the submission-level predictions, but they did not provide significant gains. Therefore, we investigated the application of a multiple linear regression model using the per-part predictions as predictors for each individual grading system and for four combinations of them. The results on LinGen and LinBus are reported in Table \ref{T:summary_results_linear_per_part}. The combinations show performances that are aligned to or better than the individual models across all metrics, with the combination including all three grading systems achieving the best results. This combination also overcomes the issue of the wav2vec2-based grader related to scoring higher levels, as can be seen in Figure \ref{fig:scatter_dev02_comb}. Table \ref{T:performance_linear} reports the $\beta$ coefficients of the individual models described in Table \ref{T:summary_results_linear_per_part} and the combination of all three. In the standard grader, as well as in the BERT-based grader, Parts 1, 2 and 5 affect the linear model most, whereas in the wav2vec2-based grader Part 1 and 5 are the most influential. In their combination, it appears that the highest $\beta$ coefficients are Parts 1 and 5 of the wav2vec2-based grader and Part 2 of the BERT-based and the standard graders. These values seem to confirm the RMSE results shown in Table \ref{T:lingen_rmse}.

\begin{table}[htb!]
\footnotesize
\begin{center}
\begin{tabular}{cccccccc}
\hline
\multicolumn{2}{c}{\textbf{Model}} & \textbf{P1} & \textbf{P2} & \textbf{P3} & \textbf{P4} & \textbf{P5} & $\beta_0$\\
\hline
\multicolumn{2}{c}{${\tt s^\otimes_{d}}$} & 0.23 & 0.25 & 0.14 & 0.15 & 0.22 & -0.11 \\
\multicolumn{2}{c}{${\tt b^\otimes_{t}}$} & 0.20 & 0.26 & 0.13 & 0.17 & 0.23 & -0.13\\
\multicolumn{2}{c}{${\tt w^\otimes_{v}}$} & 0.29 & 0.05 & 0.01 & 0.01 & 0.45 & 0.76\\
\hline\multirow{3}*{${\tt s_d}$$\otimes$${\tt b_t}$$\otimes$${\tt w_v}$} & ${\tt s_d}$ & -0.01 & 0.12 & 0.06 & 0.01 & -0.04 & \multirow{3}*{0.20}\\

& ${\tt b_{t}}$ & 0.06 & 0.16 & 0.05 & 0.09 & 0.09 & \\
& ${\tt w_v}$ & 0.20 & -0.08 & -0.02 & 0.02 & 0.20 & \\
\hline

\end{tabular}
\end{center}
\caption{$\beta$ coefficients of per-part linear regression model for the standard (${\tt s^\otimes_{d}}$), BERT (${\tt b^\otimes_{t}}$), wav2vec2 (${\tt w^\otimes_{v}}$), and combination (${\tt s_{d}}$$\otimes$${\tt b_{t}}$$\otimes$${\tt w_{v}}$) estimated on the calibration data.}

\label{T:performance_linear}

\end{table}


\begin{table}[ht!]
\small
\begin{center}

\begin{tabular}{ l  c  c  c  c  c  } \hline
\multicolumn{1}{c}{}  &     \multicolumn{5}{c}{\textbf{LinGen}} \\
 \textbf{Model} & \textbf{PCC} & \textbf{SRC} & \textbf{RMSE} & \textbf{\%$\leq$0.5} & \textbf{\%$\leq$1.0}  \\ \hline 

 ${\tt s^\otimes_d}$ & 0.932   & 0.937   & 0.382   & 82.3   & 98.7  \\
 ${\tt b^\otimes_t}$ & 0.930   & 0.935   & 0.393   & 80.3   & 98.6   \\
 ${\tt w^\otimes_v}$ & 0.933   & 0.937   & 0.393   & 79.7   & 99.0   \\
\hline
\hline
 ${\tt s_d}$$\otimes$${\tt w_v}$ & 0.941     & 0.945     & 0.363      & 84.5     & 99.3     \\

 ${\tt s_d}$$\otimes$${\tt b_t}$ & 0.936     & 0.940     & 0.373      & 81.9     & 98.8     \\

 ${\tt b_t}$$\otimes$${\tt w_v}$ & 0.943     & 0.947     & 0.359      & 84.3     & 99.2     \\
\hline
\hline
\hline
 ${\tt s_d}$$\otimes$${\tt b_t}$$\otimes$${\tt w_v}$ & 0.943     & 0.947     & 0.356      & 85.0     & 99.1     \\
\hline

\end{tabular}
\end{center}

\begin{center}

\begin{tabular}{ l  c  c  c  c  c } \hline
\multicolumn{1}{c}{} &     \multicolumn{5}{c}{\textbf{LinBus}} \\
  \textbf{Model} & \textbf{PCC} & \textbf{SRC} & \textbf{RMSE} & \textbf{\%$\leq$0.5} & \textbf{\%$\leq$1.0}  \\ \hline 

${\tt s^\otimes_d}$ & 0.912  & 0.920 & 0.415   & 77.0  & 99.0    \\
${\tt b^\otimes_t}$ & 0.920   & 0.924   & 0.400   & 80.1   & 99.0   \\
${\tt w^\otimes_v}$ & 0.916  & 0.919  & 0.394   & 79.1   & 99.0     \\
\hline
\hline
 ${\tt s_d}$$\otimes$${\tt w_v}$ & 0.925     & 0.928    &  0.378    & 82.0     & 99.4     \\

 ${\tt s_d}$$\otimes$${\tt b_t}$ & 0.925     & 0.929     & 0.391      & 80.8     & 99.4     \\

 ${\tt b_t}$$\otimes$${\tt w_v}$ & 0.930     & 0.932     & 0.368      & 82.7     & 99.3     \\
 \hline
 \hline
 \hline
 ${\tt s_d}$$\otimes$${\tt b_t}$$\otimes$${\tt w_v}$ & 0.931     & 0.933     & 0.366      & 82.5     & 99.4     \\

\hline

\end{tabular}
\end{center}

\caption{Results on overall grades on LinGen and LinBus using per-part linear regression estimated on the calibration data.}
\label{T:summary_results_linear_per_part}
\end{table}

\section{CONCLUSIONS}

In this study, we have extended our recent novel approach to proficiency assessment using a wav2vec2-based grader applying it to a large quantity of L2 learner data. 
First, we compared its performance on the five parts of the Linguaskill exam to a high performing standard grader fed with hand-crafted features and a BERT-based grader. 
We found that our proposed approach appears to be sensitive to the nature of the responses for a part with good performance for parts consisting of short spontaneous answers. 
Secondly, we found that the three grading systems have comparable performances on overall grades, with the wav2vec2-based grader showing some difficulties in assessing higher levels. Finally, we combined the standard and the wav2vec2-based graders by means of various linear combinations and we found interesting improvements. 
A concern with the wav2vec2 and BERT-based graders is that they are not fully valid alone since neither considers all aspects of the assessment construct and their results are not interpretable to provide feedback to a learner. As well as boosting the assessment performance, combination with the standard hand-crafted feature grader removes these concerns. 



\setstretch{0.85}

\bibliographystyle{IEEEbib}
\bibliography{refs}

\end{document}